\documentclass[%
 reprint,
 amsmath,amssymb,
 aps,
]{revtex4}

\usepackage{graphicx}% Include figure files
\usepackage{dcolumn}% Align table columns on decimal point
\usepackage{bm}% bold math
\usepackage{subfigure}
\begin{document}

\preprint{APS/123-QED}

\title{Beta-function formalism for k-essence constant-roll inflation}% Force line breaks with \\
%\author{NAME NAME}
\author{Abolhassan Mohammadi$^{a,b}$}
 \email{abolhassanm@gmail.com;a.mohammadi@uok.ac.ir}
\author{Tayeb Golanbari$^a$}%
 \email{t.golanbari@gmail.com}
 \author{Khaled Saaidi$^a$}%
 \email{ksaaidi@uok.ac.ir}
\affiliation{%
 $^a$Department of Physics, Faculty of Science, University of Kurdistan, Iran. \\
$^b$Dipartimento di Fisica e Astronomia, Universita di Bologna, via Irnerio 46, I-40126 Bologna, Italy.
}%


%\affiliation{
%\affiliation{
%\affiliation{%


\date{\today}% It is always \today, today,

\begin{abstract}
The beta-function formalism is being used to study the constant-roll inflation in k-essence models. Assuming the second slow-roll parameter as a constant leads to a first order differential equation for $\beta$-function which much easier to solve and find a solution than the second (non-liner) order equation that we have in the corresponding standard constant-roll inflation. Many cosmological models are known as a subclass of k-essence, so we will try to consider the model as general as possible. It is determined that the second slow-roll parameter should be positive to produce a small value for the first slow-roll parameter. The scenario is considered for three well-known cosmological model, and it is clarified that for non-canonical and tachyon model, the scalar spectral index never reach the observational range, however for DBI model we could arrives as a result consistence with the observational data.
%\begin{description}
%You may use the \texttt{description} environment to structure your %abstract;
%use the optional argument of the \verb+\item+ command to give the %category of each item.
%\end{description}

\end{abstract}

\maketitle

%\tableofcontents

%%%%%%%%%%%%%%%%%%%%%%%%%%%%%%%%%%%%%%%%
%%%%%%%%%%%%%%%%%%%%%%%%%%%%%%%%%%%%%
%%%%%%%%%%%%%%%%%%%%%%%%%%%%%%%%%%%%%%%%%%%%%
%%%%%%%%%%%%%%%%%%%%%%%%%%%%%%%%%%%%%%%%%%%%%
%%%%%%%%%%%%%%%%%%%%%%%%%%%%%%%%%%%%%%%%
%%%%%%%%%%%%%%%%%%%%%%%%%%%%%%%%%%%%%
%%%%%%%%%%%%%%%%%%%%%%%%%%%%%%%%%%%%%%%%%%%%%
\section{Introduction}\label{intro}
The inflationary scenario is one of the best candidate to describe the evolution of the universe in very early times. The first type of inflationary model was proposed by Starobinsky \cite{starobinsky1980new} which was based on the conformal anomaly in quantum gravity. The model was very complicated and the main goal was not to solve the homogeneity and isotropy problems. As a matter of fact, it was assumed that the universe is homogeneity and isotropy from the very beginning. A simpler model of inflation was introduced by A. Guth \cite{Guth:1980zm}, with this aim to solve the drawbacks of the hot big-bang model. Although the model, which is known as old inflation, did not work properly, the idea was so elegant that it had a deep impact on future inflationary models. \\
In a years after proposing the idea, it became clear that the scenario has its own defects, but using the idea of the expansion the scenario of new inflation was propose in two independent works \cite{albrecht1982cosmology,linde1982new}. And finally in 1983, Linde \cite{linde1983chaotic} propose the chaotic inflation which stands mostly on slow-rolling assumption. So far, and by relying on slow-roll assumptions, many inflationary models have been proposed such as non-canonical inflation \cite{Barenboim:2007ii,Franche:2010yj,Unnikrishnan:2012zu,Gwyn:2012ey,Rezazadeh:2014fwa,Cespedes:2015jga,Stein:2016jja,Pinhero:2017lni,Teimoori:2017wbx}, tachyon inflation \cite{Fairbairn:2002yp,Mukohyama:2002cn,Feinstein:2002aj,Padmanabhan:2002cp}, G-inflation \cite{maeda2013stability,abolhasani2014primordial,alexander2015dynamics,tirandari2017anisotropic}, brane inflation \cite{maartens2000chaotic,golanbari2014brane}, warm inflation \cite{berera1995warm,berera2000warm,taylor2000perturbation,hall2004scalar,BasteroGil:2004tg,Sayar:2017pam,Akhtari:2017mxc} and so on. Today, there is a large amount of observational data which clear our viewpoint about the earliest times of the universe evolution \cite{Planck:2013jfk,Ade:2015lrj,Akrami:2019izv}. \\
An extreme part of research on inflation is based on introducing a type scalar field potential which is mostly motivated from fundamental theory of physics such as quantum field theory, supergravity, string theory, etc. There are some other formalisms for studying inflation and the most recent formalism is addressed as $\beta$-function formalism, introduced by \cite{Binetruy:2014zya}. The motivation for this formalism comes from an analogy between the dynamical equation of the scalar field in FLRW background and a renormalization group equation. The similarity of the equations motivated the authors of \cite{Binetruy:2014zya} to introduced a $\beta$-function for the evolution of the universe; the same process that is performed in quantum field theory. They found that as $\beta$-function becomes smaller than unity, the universe experiences an accelerated expansion phase, and for $\beta=0$ there is an exact de-Sitter universe. So it is expected that the inflation occurs as the $\beta$-function is near to zero. \\

Slow-roll inflationary scenario relies on this assumption that the slow-roll parameters during the inflation are smaller than unity and the second slow-roll parameter $\eta={\ddot\phi \over H\dot\phi}$ includes to this fact. What happens if this parameter is exactly zero? This question first was considered by \cite{Kinney:2005vj}, where it was shown that in this situation the curvature perturbations could evolve on the superhorizon scales. In \cite{Namjoo:2012aa}, the non-Gaussianity of such a model was considered and the result determines that in this case the non-Gassuanity could be of order one in contrast to the standard slow-roll inflation where the non-Gaussianity parameter is very small. The authors of \cite{Martin:2012pe} generalized the case a little more and assumes that the slow-roll parameter $\eta$ is a constant, and they could find an approximate solution for the model. Then, by considering the scalar perturbations it was shown that, in general, the amplitude of the scalar perturbation could be evolve even on superhorizon scales. The work has been improved in \cite{Motohashi:2014ppa}, where the authors use Hamilton-Jacobi formalism \cite{Salopek:1992qy,Liddle:1994dx,Kinney:1997ne,Guo:2003zf,Aghamohammadi:2014aca,Saaidi:2015kaa,Sheikhahmadi:2016wyz} and extract an exact solution for the model. It was determined that the solution confirms attractor behavior. Also, investigating the cosmological perturbation for the model showed that the amplitude of curvature perturbations evolves on superhorizon scale, and to have an almost scale invariant feature, the second slow-roll parameter should be chosen properly by using observational data. It worth to mention that the word "constant-roll inflation" was first addressed in \cite{Motohashi:2014ppa}. The scenario has received cosmologists interest and many research have been performed regarding the topic. \\
During this work, we are about to use the $\beta$-function formalism for studying constant-roll inflation. The k-essence model is a wide cosmological model which many other cosmological model such as tachyon and DBI are subclass of this model. The power of the $\beta$-formalism in considering the inflationary scenario motivates us to use this formalism in constant-roll inflation as well. The canonical model of constant-roll inflation with $\beta$-function formalism has been studied in \cite{Cicciarella:2017nls}, and it is determined that there could be a first order differential equation for the $\beta$-function instead of the second order differential equation for the Hubble parameter [??]. By deriving the solutions of the model, it is found out that they could agree with observational data. Here, the k-essence model, in the general form, is used to describe constant-roll inflation. Applying $\beta$-function formalism leads to a first order differential equation; in contrast to the Hamilton-Jacobi formalism which gives a second order differential equation. It is tried to derive the parameters as general as possible so that by extracting the $\beta$-function from the differential equation, one could obtain the Hubble parameter, time derivative of the scalar field, the slow-roll parameters and also the perturbation parameters such as amplitude of the scalar perturbation, scalar spectral index and tensor-to-scalar ratio in a general form.  \\
The paper is organized as follows: In Sec.\ref{kessencemodel}, we briefly review the k-essence model and present the main dynamical equations. Then, the $\beta$-function is introduced. The slow-roll parameters are introduced in Sec.\ref{seciii:constantroll}, and they are expressed in terms of the $\beta$, and by taking the second slow-roll parameter as a constant, a first order differential equation is obtained for $\beta$ which leads to a exact solution for the model. Sec.\ref{seciv:perturbation} dedicated to the perturbation parameters and they are expressed in general form. Comparing the results of the model with observational data is performed in Sec.\ref{secv:result&observation}, where the main perturbation parameters are estimated and the proper values of the model constant are determined. The results will be sum up in Sec.\ref{conclusion}.

%%%%%%%%%%%%%%%%%%%%%%%%%%%%%%%%%%%%%%%%
%%%%%%%%%%%%%%%%%%%%%%%%%%%%%%%%%%%%%
%%%%%%%%%%%%%%%%%%%%%%%%%%%%%%%%%%%%%%%%%%%%%
%%%%%%%%%%%%%%%%%%%%%%%%%%%%%%%%%%%%%%%%%%%%%
%%%%%%%%%%%%%%%%%%%%%%%%%%%%%%%%%%%%%%%%
%%%%%%%%%%%%%%%%%%%%%%%%%%%%%%%%%%%%%
%%%%%%%%%%%%%%%%%%%%%%%%%%%%%%%%%%%%%%%%%%%%%
\section{k-essence model and $\beta$-function}\label{kessencemodel}
k-eesence model is a very well known model in cosmological studies. Many outstanding cosmological models are subclass of this model such as non-canonical, Tachyon and DBI model. The action of the model in a general form is given as
\begin{equation}\label{action}
S = \int d^4x \sqrt{-g} \left( {1 \over 2} \; R + \mathcal{L}(X,\phi) \right)
\end{equation}
where $g$ is the determinant of the metric $g_{\mu\nu}$, and $R$ is the scalar curvature constructed from the metric. $\phi$ stands for the scalar field and $X$ is defined as $X=g^{\mu\nu} \nabla_\mu \phi \nabla_\nu \phi /2$. The Lagrangian of the scalar field is indicated by $\mathcal{L}(X,\phi)$ which in general is an arbitrary function of $X$ and the scalar field $\phi$. Variation comes to the same field equation $G_{\mu\nu} = T_{\mu\nu}$ and the only difference is in the component of the scalar field energy momentum tensor so that for a homogeneous and isotropic universe there are
\begin{equation}\label{energypressure}
    \rho = 2X \mathcal{L}_{,X} - \mathcal{L}, \qquad
    p = \mathcal{L}.
\end{equation}
and sound speed, indicated by $c_s$, is given by
\begin{equation}\label{cs}
    c_s^2 = {p_{,X} \over \rho_{,X}} = {\mathcal{L}_{,X} \over \mathcal{L}_{,X} + 2X\mathcal{L}_{,XX}}.
\end{equation}
which the exact form depends on the form of the Lagrangian. Assuming a homogeneous and isotropic  space-time we are following the spatially flat FLRW metric as
\begin{equation}\label{metric}
    ds^2 = -dt^2 + a^2(t) \; \left( dx^2 + dy^2 + dz^2 \right).
\end{equation}
and by applying it in the field equations, the evolution equations come to
\begin{equation}\label{friedmann}
H^2 = 2X \mathcal{L}_{,X} - \mathcal{L}, \qquad
\dot{H} = - X \mathcal{L}_{,X}
\end{equation}
In the Hamilton-Jacobi formalism, the Hubble parameter is introduced as a function of the scalar field which is taken as clock of the system $\phi(t)$. Here, it will be useful to replace the Hubble parameter with the superpotential as $W(\phi) = -2H(\phi)$, so the Friedmann equations (\ref{friedmann}) are rewritten as
\begin{equation}\label{Wfriedmann}
   {3 \over 4} W^2(\phi) = 2X \mathcal{L}_{,X} - \mathcal{L}, \qquad W'(\phi) = \mathcal{L}_{,X} \dot{\phi}
\end{equation}
where prime denotes derivative with respect to the scalar field.  Eq.\eqref{Wfriedmann} is the Hamilton-Jacobi equation of the model, in which by imposing an specific scalar field Lagrangian, the potential of the model is expressed in terms of the superpotential $W(\phi)$ and its derivative. For instance, for three well-known k-essence model, one has
\begin{enumerate}
  \item For non-canonical scalar field model with Lagrangian $\mathcal{L}(X,\phi)= X \left( X / M \right)^{\alpha-1} - V(\phi)$, the Friedmann equations are given by
      \begin{equation}\label{non-friedmann}
        3H^2 = (2\alpha -1)X \left( X^\alpha \over M \right)^{\alpha-1} + V(\phi), \qquad -2\dot{H} = 2\alpha \; X \left( X^\alpha \over M \right)^{\alpha-1}
      \end{equation}
      the potential of scalar field could be obtained from the Firedmann equations as
     \begin{equation}\label{noncpot}
     V(\phi) = {3 \over 4} \; W^2(\phi) - {(2\alpha-1) \over 2^\alpha M^{4(\alpha-1)}} \left[ {2^{\alpha-1} M^{4(\alpha-1)} \over \alpha} \; W'(\phi) \right]^{2\alpha \over 2\alpha-1},
     \end{equation}
     where the identity $\dot{H}= \dot{\phi} H'$ has been utilized. The sound speed of the model is acquired from Eq.\eqref{cs} and by using the Lagrangian  as
     \begin{equation}\label{csnon}
      c_s = {1 \over \sqrt{2\alpha -1}},
     \end{equation}
     which is a constant and for $\alpha=1$ comes back to the canonical model of scalar field, i.e. $c_s=1$. \\

  \item For the tachyon scalar field model the Lagrangian is given by $\mathcal{L}(X,\phi)= -V(\phi) \sqrt{1-2X}$, which leads to the following Friedmann equations
      \begin{equation}\label{tac-friedmann}
        3H^2 = {V(\phi) \over \sqrt{1 - \dot{\phi}}}, \qquad -2\dot{H} = 3 H^2 \dot{\phi}^2
      \end{equation}
      From above equations and by using the fact that the time derivative of the Hubble parameter could be written as $\dot{H} = \dot{\phi} H'$, the potential of the tachyon field is reached as
    \begin{equation}\label{tachyonpot}
     V(\phi) = {3 \over 4} \; W^2(\phi) \; \sqrt{1 - {8 \over 9} \; {W^{\prime 2}(\phi) \over W^4(\phi)}}
    \end{equation}
    where we have applied the superpotential $W(\phi)$ instead of the Hubble parameter. Based on the definition \eqref{cs}, the sound speed for the model is read as
    \begin{equation}\label{csnon}
     c_s = \sqrt{1 + {2 \over 3} \; {\dot{H} \over H^2}} = = \sqrt{1 - {2 \over 3} \; \epsilon }
    \end{equation}
    where $\epsilon = -\dot{H / H^2}$ is known as the first slow-roll parameter which will be introduced in next section. \\

  \item For the DBI scalar field model, where the Lagrangian is $\mathcal{L}(X,\phi)= -f^{-1}(\phi) \left[ \sqrt{1-2f(\phi)X} - 1 \right]-V(\phi)$, one arrives at the following Friedmann equations
      \begin{equation}\label{dbifriedmann}
      3H^2 = {\gamma - 1 \over f(\phi)} + V(\phi) , \qquad -2\dot{H} = \gamma \dot{\phi}^2,
      \end{equation}
      in which the parameter $\gamma$ is known as the Lorentz coefficient and defined as
      \begin{equation}\label{gamma}
        \gamma = {1 \over \sqrt{1 - f(\phi) \dot{\phi}^2}}.
      \end{equation}
      From the second Friedmann equation of the DBI model \eqref{dbifriedmann} and by notting Eq.\eqref{Wfriedmann}, it is found out that the $\gamma$ parameter is the derivative of the Lagrangian of the model, i.e.
      \begin{equation}\label{gammaLx}
        \gamma = \mathcal{L}_{,X}
      \end{equation}
      which also could be verified by taking derivative of the Lagrangian. \\
      Working with the Friedmann equations, the potential for DBI scalar field is obtained as
      \begin{equation}\label{dbipot}
      V(\phi) = {3 \over 4} \; W^2(\phi) - { (\gamma - 1) \over f(\phi)}
      \end{equation}
      From the definition of the $\gamma$ parameter and by using the second Friedmann equation, the function $f(\phi)$, which is addressed as the brane tension, could be written as
      \begin{equation}\label{fphi}
      f(\phi) = {\gamma^2 - 1 \over W^{\prime 2}(\phi)} .
     \end{equation}
     Then, the potantial \eqref{dbipot} is rewritten as
     \begin{equation}\label{dbipotential}
      V(\phi) = {3 \over 4} \; W^2(\phi) - { W^{\prime 2} \over \gamma + 1}
                 = {3 \over 4} \; W^2(\phi) \left[ \; 1 - {2 \over 3} \; {\gamma \; \epsilon \over \gamma + 1} \; \right].
      \end{equation}
      From the definition of sound speed $\eqref{cs}$, it is realized that the sound speed for the model is related to the inverse of the $\gamma$ parameter
      the potential and sound speed are given by
     \begin{equation}\label{dbics}
     c_s = {1 \over \gamma} = \sqrt{1 - f(\phi) \dot{\phi}^2}
     \end{equation}

\end{enumerate}

According to \cite{Binetruy:2016hna,Pieroni:2015cma}, the $\beta$-function is introduced as
\begin{equation}\label{beta}
    \beta(\phi) = \sqrt{\mathcal{L}_{,X}} \; {d\phi \over d\ln(a)} = \sqrt{\mathcal{L}_{,X}} \; {\dot\phi \over H} = {-2 \over \sqrt{\mathcal{L}_{,X}} } \; {W'(\phi) \over W(\phi)}.
\end{equation}
By this definition, it could be easily confirmed that the equation of state is found out only in terms of $\beta$ in which
\begin{equation}\label{EoS}
    {\rho + p \over \rho} = {1 \over 3} \; {\beta^2(\phi)}
\end{equation}
Then, the inflation could happen in the vicinity of zero of $\beta$.

%%%%%%%%%%%%%%%%%%%%%%%%%%%%%%%%%%%%%%%%
%%%%%%%%%%%%%%%%%%%%%%%%%%%%%%%%%%%%%
%%%%%%%%%%%%%%%%%%%%%%%%%%%%%%%%%%%%%%%%%%%%%
%%%%%%%%%%%%%%%%%%%%%%%%%%%%%%%%%%%%%%%%%%%%%
%%%%%%%%%%%%%%%%%%%%%%%%%%%%%%%%%%%%%%%%
%%%%%%%%%%%%%%%%%%%%%%%%%%%%%%%%%%%%%
%%%%%%%%%%%%%%%%%%%%%%%%%%%%%%%%%%%%%%%%%%%%%
\section{SRP and constant-roll inflation}\label{seciii:constantroll}
The inflation is known as a phase of accelerated expansion for the earliest times, where the universe undergoes an extreme expansion in short period of time. The inflation usually describe through some slow-roll parameters, for instance the first one is $\epsilon=-\dot{H}/H^2$ which is the variation of the Hubble parameter in a Hubble time. If this parameter is smaller than one, $\epsilon<1$, the acceleration phase of the universe is guaranteed so that $\ddot{a}/a = H^2 (1-\epsilon)$. The other slow-roll parameters are defined through a hierarchy treatment as if the first one is
\begin{equation}
    \epsilon_1 = - {\dot{H} \over H^2}
\end{equation}
the next ones are given as
\begin{equation}\label{srp}
    \epsilon_{i+1} = {\dot{\epsilon}_i \over H \epsilon_i}
\end{equation}
Now we are bout to rewrite the slow-roll parameters in terms of $\beta$. Using Eqs.(\ref{friedmann}) and (\ref{EoS}), the first slow-roll parameters could be expressed as
\begin{equation}\label{epsilon}
\epsilon_1 = -{\dot{H} \over H^2} = {3 \over 2} \; {\rho + p \over \rho} = {1 \over 2} \; \beta^2(\phi),
\end{equation}
which received a very simple form. The second slow-roll parameter is given through the definition (\ref{srp}) and Eq.(\ref{beta}) as
\begin{equation}\label{epsilon2}
    \epsilon_2 = {\dot{\epsilon}_1 \over H \epsilon_1} = {2 \beta'(\phi) \over \sqrt{\mathcal{L}_{,X}}}.
\end{equation}
The first slow-roll parameter should be smaller than unity in constant-roll inflationary scenario as well. The story of the second slow-roll parameter is different. In constant-roll inflationary scenario, the second slow-roll parameters is assumed as a constant which in general it might be of order unity and not small. Taking into account this assumption, $\epsilon_2=\lambda$, one arrives at the following differential equation for $\beta$
\begin{equation}\label{betadifeq}
    \beta'(\phi) = {\lambda \over 2} \; \sqrt{\mathcal{L}_{,X}}.
\end{equation}
Here in k-essence model $\mathcal{L}_{,X}$ could be in general a function of both scalar field $\phi$ and $\dot\phi$. By applying this differential equation on Eq.(\ref{beta}), the superpotential could be exported as a function of scalar field as
\begin{equation}\label{W}
  W(\phi) = W_0 \exp\left( {-1 \over 2\lambda} \; \beta^2(\phi) \right),
\end{equation}
where $W_0$ is a constant of integration. From the definition, we have $2H(\phi) = -W(\phi)$. Since the Hubble parameter is a positive quantity during the inflation, the function $W(\phi)$ and then $W_0$ must be negative.  \\
Solving the first order differential equation (\ref{betadifeq}), and getting an exact solution for $\beta$, first need to specify the term $\mathcal{L}_{,X}$ as a function of the scalar field. \\
The amount of expansion during inflation is given by the number of e-fold parameter which is obtained as
\begin{equation}
    N = \int^{t_e}_{t_\star} H dt = \int^{\phi_e}_{\phi_\star} {H \over \dot\phi} d\phi = \int^{\phi_e}_{\phi_\star} {\sqrt{\mathcal{L}_{,X}} \over \beta(\phi)} \; d\phi = \int^{\phi_e}_{\phi_\star} {\sqrt{\mathcal{L}_{,X}} \over \beta(\phi) \beta'(\phi)} \; d\beta = {2 \over \lambda} \int^{\phi_e}_{\phi_\star} { d\beta \over \beta(\phi)},
\end{equation}
where Eq.(\ref{betadifeq}) has been used, and here the subscripts $"e"$ and $"\star"$ respectively stands for the end of inflation and the time of horizon crossing. Taking integrate leads one to
\begin{equation}\label{efoldbeta}
  \beta(\phi_\star) = \beta(\phi_e) \; e^{-\lambda N \over 2}.
\end{equation}
In addition, at the end of inflation
\begin{equation}\label{betaepsilonend}
 \epsilon(\phi_e) = \beta^2(\phi_e)/2=1,
\end{equation}
then, in general, there is
\begin{equation}\label{betaepsilonstar}
\beta^{\star} = \sqrt{2} \; e^{-\lambda N \over 2}; \qquad \epsilon_1^{\star} = e^{-\lambda N}.
\end{equation}
The fundamental assumption is that the slow-roll parameter $\epsilon$ is smaller than one at the time of horizon crossing. Due to this fact, the constant $\lambda$ must be positive otherwise the slow-roll parameter goes below one. \\
From Eq.(\ref{W}) and by imposing Eq.(\ref{betaepsilonstar}), the superpotential could be found out during inflation as a function of number of e-fold as
\begin{equation}\label{WN}
  W^\star = W_0 \exp\left( {-1 \over \lambda} \; e^{-\lambda N} \right), \qquad H^\star = {- W^\star \over 2}
\end{equation}
By determining $\beta$ in terms of the scalar field, which in turn needs to specify $\sqrt{\mathcal{L}_{,X}}$ as a function of the scalar field, the scalar field at the time of horizon crossing could also be obtained in terms of the number of e-fold.\\
In this regards, we are going to assume three typical function of the scalar field for $\sqrt{\mathcal{L}_{,X}}$ and consider the solution of the model in detail. In the following lines we introduce three common functions as power-law, exponential and hyperbolic function of $\beta(\phi)$ for the term $\sqrt{\mathcal{L}_{,X}}$ and consider the solutions.

%%%%%%%%%%%%%%%%%%%%%%%%%%%%%%%%%%%%
%%%%%%%%%%%%%%%%%%%%%%%%%%%%%%%%%%%%
%%%%%%%%%%%%%%%%%%%%%%%%%%%%%%%%%%%%
%%%%%%%%%%%%%%%%%%%%%%%%%%%%%%%%%%%%

\subsection{First Ansantz: $\sqrt{\mathcal{L}_{,X}} = A_1 \beta^n(\phi)$}
Applying this ansantz to Eq.(\ref{betadifeq}), the function $\beta$ is obtained as
\begin{equation}\label{firstbeta}
        \beta(\phi) = \left( { \lambda A_1 (1-n) \over 2}\; (\phi-\phi_0)  \right)^{1 \over 1-n}
\end{equation}
and from the above definition, we have
\begin{equation}\label{Lx1}
        \mathcal{L}_{,X} = A_1^2 \left( { \lambda A_1 (1-n) \over 2} \; (\phi - \phi_0) \right)^{2n \over 1-n}
\end{equation}
Using Eq.(\ref{beta}), the superpotential is obtained by taking an integrate which comes to
\begin{equation}\label{W1}
W(\phi) = W_0 \exp\left( {-1 \over 2\lambda} \; \left( { \lambda A_1 (1-n) \over 2}\; (\phi-\phi_0)  \right)^{2 \over 1-n} \right),
\end{equation}
and the time derivative of the scalar field during inflation could be found out by using Eq.(\ref{beta}) as
\begin{equation}\label{phidot1}
    \dot\phi = {-W_0 \over \lambda A_1^2 (1-n) \; (\phi-\phi_0)} \; \exp\left[ {-1 \over 2\lambda} \; \left( { \lambda A_1 (1-n) \over 2}\; (\phi-\phi_0)  \right)^{2 \over 1-n} \right].
\end{equation}
The first slow-roll parameter $\epsilon_1$ is given in Eq.(\ref{epsilon}), and is read as
\begin{equation}\label{epsilon01}
\epsilon_1(\phi) = {1 \over 2} \; \left( {(1-n) \lambda A_1 \over 2} \; (\phi - \phi_0) \right)^{2 \over 1-n}
\end{equation}
and by substituting Eqs.(\ref{betaepsilonend}) and (\ref{betaepsilonstar}) into (\ref{firstbeta}), the scalar field at the end and horizon crossing time of inflation are respectively given by
\begin{eqnarray}
 (\phi_e-\phi_0) &=& { 2^{3-n \over 2} \over (1-n) \lambda A_1 } \nonumber \\
  (\phi_\star - \phi_0 ) &=& { 2^{3-n \over 2} \over (1-n) \lambda A_1 } \; \exp\left( {(n-1)\lambda \over 2} \; N \right).
\end{eqnarray}
By assuming that during inflation the term $(\phi-\phi_0)$ is always positive, From Eq.(\ref{phidot1}) it is clear that for $n>1$($<1$), the time derivative of the scalar field is negative (positive) and then the scalar field decreases (increases) during the time. On the other hand, since we claim for having a positive $(\phi-\phi_0)$, the constant $A_1$ should be negative (positive).

%%%%%%%%%%%%%%%%%%%%%%%%%%%%%%%%%%%%
%%%%%%%%%%%%%%%%%%%%%%%%%%%%%%%%%%%%
%%%%%%%%%%%%%%%%%%%%%%%%%%%%%%%%%%%%
%%%%%%%%%%%%%%%%%%%%%%%%%%%%%%%%%%%%
\subsection{Second Ansantz: $\sqrt{\mathcal{L}_{,X}} = A_2 \exp\left( M_2 \beta(\phi) \right)$}
For this choice, $\beta$-function is obtained from differential equation (\ref{betadifeq}) as
\begin{equation}\label{secondbeta}
    \beta(\phi) = {-1 \over M} \; \ln\left( {-M_2 A_2 \lambda \over 2} \; (\phi - \phi_0) \right),
\end{equation}
where $\phi_0$ is a constant of integration. Following the definition, the term $\mathcal{L}_{,X}$ is given by
\begin{equation}\label{Lx2}
    \mathcal{L}_{,X} = {4 \over M_2^2 \lambda^2 (\phi-\phi_0)^2},
\end{equation}
and using Eq.(\ref{beta}), the superpotential is realized as
\begin{equation}\label{W2}
    W(\phi) = W_0 \exp\left( {-1 \over 2 M_2^2 \lambda} \; \left[ \ln\left( {-A_2 M_2 \lambda \over 2} \; (\phi-\phi_0) \right). \right]^2 \right)
\end{equation}
Inserting above superpotential in Eq.(\ref{beta}) leads one to the time derivative of the scalar field which is
\begin{eqnarray}\label{phidot2}
    \dot\phi & = & {-\lambda W_0 \over 4} \; (\phi-\phi_0) \;
    \ln\left[ {-M_2 A_2 \lambda \over 2} \; (\phi-\phi_0) \right] \\
 & & \quad \exp\left[ {-1 \over \lambda M_2^2} \left( \ln\left[ {-M_2 A_2 \lambda \over 2} \; (\phi-\phi_0) \right] \right)^2 \right]. \nonumber
\end{eqnarray}
From EQ.(\ref{beta}), the first slow-roll parameter is acquired as follows
\begin{equation}
    \epsilon_1 = {1 \over 2M_2^2} \; \left[ \ln\left( {-A_2 M_2 \lambda \over 2} \; (\phi - \phi_0) \right) \right]^2.
\end{equation}
Inflation ends as the slow-roll parameter $\epsilon_1$ reaches unity. Then, using Eqs.(\ref{betaepsilonend}), (\ref{betaepsilonstar}) and (\ref{secondbeta}), one could find the scalar field at the end of inflation and at the horizon crossing time respectively as
\begin{eqnarray}\label{phiendstar2}
  \phi_e - \phi_0 &=& {-2 \over A_2 M_2 \lambda} \; \exp\left( - \sqrt{2} \; M_2 \right), \nonumber \\
  \phi_\star - \phi_0 &=& {-2 \over A_2 M_2 \lambda} \; \exp\left[ - \sqrt{2} \; M_2  \; \exp\left( {-\lambda N \over 2} \right)  \right].
\end{eqnarray}
Assuming $(\phi-\phi_0)>0$, then $A_2M_3$ should be negative to release the logarithmic term from being indefinite. Then, to have $\dot\phi > 0$, the condition ${-A_2 M_2 \lambda \over 2} \; (\phi - \phi_0) > 1$ should be satisfied, otherwise the time derivative of the scalar field will be negative. Now suppose that ${-A_2 M_2 \lambda \over 2} \; (\phi - \phi_0) > 1$, then by investigating Eq.(\ref{phiendstar2}) it is realzed that the constant $M_2$ is negative to satisfy the increasing behavior of the scalar field. And if ${-A_2 M_2 \lambda \over 2} \; (\phi - \phi_0) < 1$ which means $\dot\phi<0$, Eq.(\ref{phiendstar2}) states that the constant $M_2$ be positive to satisfy the decreasing behavior of the scalar field.

%%%%%%%%%%%%%%%%%%%%%%%%%%%%%%%%%%%%
%%%%%%%%%%%%%%%%%%%%%%%%%%%%%%%%%%%%
%%%%%%%%%%%%%%%%%%%%%%%%%%%%%%%%%%%%
%%%%%%%%%%%%%%%%%%%%%%%%%%%%%%%%%%%%
\subsection{Third Ansantz: $\mathcal{L}_{,X} = {A_3 \over \cosh\left( M_3 \beta(\phi) \right)}$}
Imposing this choice on Eq.(\ref{betadifeq}) comes to the following expression for $\beta$
\begin{equation}
        \beta(\phi) = {1 \over M_3} \; \sinh^{-1}\left( {A_3 M_3 \lambda \over 2 } \; (\phi - \phi_0) \right),
\end{equation}
where $\phi_0$ is a constant of integration. Then, for $\mathcal{L}_{,X}$ we have
\begin{equation}\label{Lx3}
\mathcal{L}_{,X} = \left( A_3 \over \cosh\left[ \sinh^{-1}\left( {A_3 M_3 \lambda \over 2 } \; (\phi - \phi_0) \right) \right] \right)^2.
\end{equation}
By determining $\beta(\phi)$, the superpotential will be clear by taking an integrate, so that it is obtained as
\begin{equation}\label{W3}
        W(\phi) = W_0 \exp\left[ {-1 \over 2 M_3^2 \lambda} \;  \left[ \sinh^{-1}\left( {A_3 M_3 \lambda \over 2 } \; (\phi - \phi_0) \right) \right]^2  \right].
\end{equation}
Then, using Eq.(\ref{Wfriedmann}), the time derivative of the scalar field is given in terms of the scalar field as
\begin{eqnarray}\label{phidot3}
\dot\phi & = & {-W_0  \over 2 A_3  M_3 } \;
{ \Big( \cosh\left[ \sinh^{-1}\left( {A_3 M_3 \lambda \over 2 } \; (\phi - \phi_0) \right) \right] \Big)^2 \;
\sinh^{-1}\left( {A_3 M_3 \lambda \over 2} \; (\phi - \phi_0) \right) \over \sqrt{1 + {A_3^2 M_3^2 \lambda^2 \over 4} \; (\phi - \phi_0)^2} }\nonumber \\
    & & \qquad \qquad \times \exp\left[ {-1 \over 2 M_3^2 \lambda} \;  \left[ \sinh^{-1}\left( {A_3 M_3 \lambda \over 2 } \; (\phi - \phi_0) \right) \right]^2  \right].
\end{eqnarray}
The first slow-roll parameter has a simple relation with $\beta$ as Eq.(\ref{beta}), and is given by
\begin{equation}\label{epsilon3}
    \epsilon_1 = {1 \over 2M_3^2} \; \left[ \sinh^{-1}\left( {A_3 M_3 \lambda \over 2 } \; (\phi - \phi_0) \right) \right]^2.
\end{equation}
Following the same approaches as the previous cases, the scalar field could be read at the end of inflation and during inflation versus $N$ as
\begin{eqnarray}\label{phiendstar3}
  \phi_e - \phi_0 &=& {2 \over A_3 M_3 \lambda} \; \sinh\left( \sqrt{2}\; M_3 \right), \nonumber \\
  \phi_\star - \phi_0 &=& {2 \over A_3 M_3 \lambda} \; \sinh\left( \sqrt{2} \; M_3 \; e^{-\lambda N \over 2} \right).
\end{eqnarray}
Assume that during the inflationary time, the term $(\phi-\phi_0)$ is always positive. In this case, from Eq.(\ref{phidot3}), it is found out that the time derivative of the scalar field is always positive as well, which results in an increasing behavior for the scalar field. The positiveness of the term $\phi-\phi_0$ should also apply to the Eq.(\ref{phiendstar3}), which clarifies that the constant $A_3$ is always positive, regardless of the sign of the constant $M_3$.

%%%%%%%%%%%%%%%%%%%%%%%%%%%%%%%%%%%%%%%%%%%%%
%%%%%%%%%%%%%%%%%%%%%%%%%%%%%%%%%%%%%%%%%%%%%
%%%%%%%%%%%%%%%%%%%%%%%%%%%%%%%%%%%%%%%%%%%%%
%%%%%%%%%%%%%%%%%%%%%%%%%%%%%%%%%%%%%%%%%%%%%
%%%%%%%%%%%%%%%%%%%%%%%%%%%%%%%%%%%%%%%%%%%%%
%%%%%%%%%%%%%%%%%%%%%%%%%%%%%%%%%%%%%%%%%%%%%
%%%%%%%%%%%%%%%%%%%%%%%%%%%%%%%%%%%%%%%%%%%%%
\section{Cosmological Perturbations}\label{seciv:perturbation}
An interesting feature of inflationary scenario is predicting cosmological perturbations which are divided to scalar, vector and tensor perturbations. Up to the linear approximation, these three types evolve independently. Scalar perturbations are known as the seed of large scale structure of the universe, and tensor perturbations are also addressed as gravitational waves. However, vector perturbations are proportional to the inverse of the scale factor of the universe and during inflation because of extremely growth of the universe, it drops fast and is ignored. The most important perturbations parameters are known as amplitude of curvature perturbation, scalar spectral index, and tensor-to-scalar ratio which is an indirect measure of the tensor perturbations. The latest observational data indicates that the primordial spectrum is almost scale invariant and it does not evolve after horizon crossing. There are many slow-roll inflationary scenario which are in perfect agreement with the results however the scenario of constant-roll inflationary scenario is different. In constant-roll inflationary scenario the second slow-roll parameter is taken as a constant which in general could be even of order one. This assumption allows the power spectrum to evolve even on superhorizon scales. Then, in order to have a valid model for describing universe inflation, the parameter needs to be constraint with observational data. \\
In this section we are going to obtain the main perturbations parameters of k-essence model in constant-roll inflation, which are mainly based on \cite{Garriga:1999vw}. The subject has been studied in \cite{Mohammadi:2018wfk,Mohammadi:2018oku,Mohammadi:2018zkf} in detail and the reader could refer to them for more information. Then we are going to follow the procedure in brief. An small perturbation of the scalar field induced a perturbation to the metric, in which in longitudinal gauge it could be written as
\begin{equation*}
ds^2 = \big(1 + 2\Phi(t,\mathbf{x}) \big) dt^2 - a^2(t) \big(1 - 2\Psi(t,\mathbf{x})\big)\delta_{ij} dx^i dx^j.
\end{equation*}
By assuming a diagonal tensor for the spatial part of the energy-momentum tensor, the action for the scalar perturbation is given by
\begin{equation}\label{vaction}
  S = {1 \over 2} \int  \left( v'^2 + c_s^2 v (\nabla v)^2 + {z'' \over z} v \right) d\tau d^3\mathbf{x}.
\end{equation}
in which $v \equiv z \left( \Phi + H \; {\delta\phi \over \dot\phi} \right)$, and prime denotes derivative with respect to the conformal time $\tau$, $a d\tau=dt$. The quantity $z$, known as the Mukhanov variable, is read as \cite{Garriga:1999vw,Mukhanov:1990me}
\begin{equation*}
  z^2 \equiv {a \; \sqrt{\rho+p} \over c_s H}.
\end{equation*}
and by imposing the Eq.(\ref{energypressure}) and Eq.(\ref{beta}) it could be expressed in terms of the $\beta$-function as
\begin{equation}\label{znew}
    z = {a \sqrt{\dot\phi^2 \mathcal{L}_{,X}} \over c_s H} = -\varepsilon {a \beta \over c_s},
\end{equation}
where $\varepsilon = \pm 1$. The perturbation equation for the variable $v$ is extracted from Eq.(\ref{vaction}) as follow
\begin{equation}\label{vkequation}
{d^2 \over d\tau^2} v_k(\tau) +\left( c_s^2 k^2 - {1 \over z}\;{d^2z \over d\tau^2} \right) \; v_k(\tau)=0.
\end{equation}
in which for subhorizon scales, where $c_sk \gg aH$ (i.e. $c_sk \gg z''/z$), it leads to the following asymptotic solution
\begin{equation}\label{vsolutionsub}
   v_k(\tau)={1 \over \sqrt{2c_s k}} e^{ic_s k \tau}.
\end{equation}
Defining the new variable $f_k$ as $v_k=\sqrt{-\tau} f_k$, the differential equation comes to the Bessel's differential equation
\begin{equation}\label{bessel}
  {d^2 f_k \over dx^2} + {1 \over x}{d f_k \over dx} + \Big( 1 - {\nu^2 \over x^2} \Big)f_k=0, \qquad
             {z'' \over z} = {\nu^2 - {1 \over 4} \over \tau^2 },
\end{equation}
where the general solution of the above differential equation clearly are the first and second kinds of Hankel function. However, by imposing the asymptotic solution for subhorizon scale, only the first kind of Hankel function remains. The spectrum of curvature perturbation is defined through the following relation
\begin{equation*}
  \mathcal{P}_s = {k^3 \over 2\pi^2}\; \left| \zeta \right|^2 = {k^3 \over 2\pi^2}\; \left| {v_k \over z} \right|^2.
\end{equation*}
in which for superhorizon scale it comes to
\begin{equation}\label{psspectrum}
  \mathcal{P}_s = \left( 2^{\nu-{3 \over 2}} \Gamma(\nu) \over \Gamma(3/2) \right)^2 \; \left( H^2 \over 2\pi \sqrt{c_s(\rho+p)} \right)^2 \;  \left( c_s k \over a H \right)^{3-2\nu}.
\end{equation}
Utilizing Eqs.(\ref{EoS}) and (\ref{epsilon}), the amplitude of the curvature perturbations at the time of horizon crossing is given as
\begin{equation}\label{ps}
    \mathcal{P}_s = {1 \over 8\pi^2} \; \left( {2^{\nu-{3 \over 2}}} \Gamma(\nu) \over \Gamma(3/2) \right)^2 {H^2 \over c_s \epsilon_1}.
\end{equation}
The scalar spectral index is given in terms of the parameter $\nu$ as
\begin{equation}\label{ns}
    n_s - 1 = 3 - 2\nu.
\end{equation}
Then, to obtain the scalar spectral index, one has to derive the derivative of $z$ with respect to the conformal time $\tau$. Before that, we are going to rewrite the parameter $z$ by using Eqs.(\ref{Wfriedmann}) and (\ref{beta}) as
\begin{equation}\label{znew}
    z = {a \sqrt{\dot\phi^2 \mathcal{L}_{,X}} \over c_s H} = -\varepsilon {a \beta \over c_s},
\end{equation}
where $\varepsilon = \pm 1$. Takin derivatives, one arrives to
\begin{equation}\label{zdd}
    {1 \over z} \; {d^2z \over d\tau^2} = (aH)^2 \; \left( 2 - \epsilon_1 + {3 \over 2} \; \lambda - 3s + {1 \over 2} \; \lambda \epsilon_1 + {1 \over 4} \; \lambda^2 - \lambda s - {\dot{s} \over H} \right),
\end{equation}
in which $s$ is another slow-roll parameters which determines the rate of sound speed in a Hubble time, i.e. $s = {\dot{c}_s \over c_s H}$. Since in quasi-de Sitter one has $a H = -(1+\epsilon_2) / \tau$, up to the first order, the parameter $\nu$ is given by
\begin{equation}\label{nu}
    \nu^2 = {9 \over 4} + 3 \epsilon_1 + {3 \over 2} \; \lambda - 3s + {5 \over 2} \; \lambda \epsilon_1 + {1 \over 4} \; \lambda^2 + {1 \over 2} \lambda^2 \epsilon_1 - \lambda s - {\dot{s} \over H}.
\end{equation}
Then, the scalar spectral index (\ref{ns}) is clear in terms of the slow-roll parameters of the model. \\

One the other hand, the second slow-roll parameter has no roll in the dynamical equation of tensor perturbations [?], and there the same relation for amplitude of tensor perturbation as we have in slow-roll inflation so that
\begin{equation}\label{pt}
    \mathcal{P}_t = {2 \; H^2 \over \pi^2}.
\end{equation}
Then using Eqs.(\ref{ps}) and (\ref{pt}), the tensor-to-scalar ratio is obtained easily
\begin{equation}
r = {\mathcal{P}_t \over \mathcal{P}_s} = 16 \left( { \Gamma(3/2)  \over 2^{\nu-{3 \over 2}} \Gamma(\nu) } \right)^2 c_s \epsilon_1 .
\end{equation}
In the following section we are going to derive the results for specific models of k-essence as non-canonical, Tachyon and DBI scalar field models and compare them with observational data and realize that what values of second slow-roll parameters are allowed by data. \\

And finally, the main perturbation parameters at the time of horizon crossing are obtained as
\begin{eqnarray}\label{perparam1}
  \mathcal{P}_s^\star &=& {1 \over 8\pi^2} \; \left( {2^{\nu^\star - {3 \over 2}} \; \Gamma(\nu^\star) \over \Gamma(3/2) } \right)^2 {W_0^2 \over c_s^\star e^{-\lambda N} } \; \exp\left( {-2 e^{\lambda N} \over \lambda} \right), \nonumber \\
  \nu^{\star 2} & = & {9 \over 4} + {3\lambda \over 2} + {\lambda^2 \over 4} + \left( 3 + {5\lambda \over 2} + {\lambda^2 \over 2} \right) \; e^{\lambda N} - \left( 3 + \lambda + {\ddot{c}_s \over c_s}\Big|_{\star} \right) \; s^{\star}, \nonumber \\
  n_s^\star &=& 4 - 2\nu^\star, \nonumber \\
  r^\star &=& 16 \left( { \Gamma(3/2) \over 2^{\nu^\star - {3 \over 2}} \; \Gamma(\nu^\star) } \right)^2 c_s^\star \; e^{-\lambda N}.
\end{eqnarray}

%%%%%%%%%%%%%%%%%%%%%%%%%%%%%%%%%%%%%%%%
%%%%%%%%%%%%%%%%%%%%%%%%%%%%%%%%%%%%%
%%%%%%%%%%%%%%%%%%%%%%%%%%%%%%%%%%%%%%%%%%%%%
%%%%%%%%%%%%%%%%%%%%%%%%%%%%%%%%%%%%%%%%%%
%%%%%%%%%%%%%%%%%%%%%%%%%%%%%%%%%%%%%%%%
%%%%%%%%%%%%%%%%%%%%%%%%%%%%%%%%%%%%%
%%%%%%%%%%%%%%%%%%%%%%%%%%%%%%%%%%%%%%%%%%%%%
\section{Results and Observational Constraints}\label{secv:result&observation}
By considering the perturbation parameters of the model in previous section, we are now in a place to study the behavior of the perturbations parameters during inflation and compare them with observational data. In this regards, an analytical solution for the function $\beta$ is required which has been done in the Sec.... for three different ansantzs, and the corresponding results will be used here when we are about to apply them at the time of horizon crossing. The results will be general until one comes to the sound speed when it is necessary to introduced an specific type of Lagrangian. The sound speed appears directly in the tensor-to-scalar ratio, and through the slow-roll parameter $s$ in the scalar spectral index. Here we are going to consider three well-known Lagrangian as non-canonical model, tachyon model and DBI model for each case. \\

%%%%%%%%%%%%%%%%%%%%%%%%%%%%%%%%%%%%%%%%
%%%%%%%%%%%%%%%%%%%%%%%%%%%%%%%%%%%%%
%%%%%%%%%%%%%%%%%%%%%%%%%%%%%%%%%%%%%%%%%%
%%%%%%%%%%%%%%%%%%%%%%%%%%%%%%%%%%%%%%%%
%%%%%%%%%%%%%%%%%%%%%%%%%%%%%%%%%%%%%
%%%%%%%%%%%%%%%%%%%%%%%%%%%%%%%%%%%%%%%%%%%%%
\subsection{Non-canonical scalar field model}
\begin{figure}[ht]
    \centering
    \includegraphics[width=6cm]{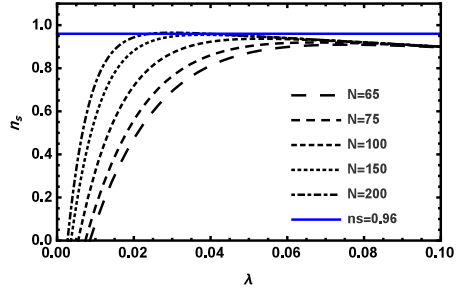}
    \caption{The scalar spectral index versus $\lambda$ for different choice of number of e-fold in the non-canonical model. }
    \label{NCnslambda}
\end{figure}

The general perturbation parameters have been determined in the previous section. Here we need to find the sound speed and its derivative, to specify the form of the scalar spectral index. The sound speed for non-canonical model is a constant
\begin{equation}\label{nonsound}
    c_s = {1 \over \sqrt{2\alpha - 1}},
\end{equation}
which leads to vanish the slow-roll parameter $s$ and its time derivative. Therefore, the parameter $\nu$ is read by
\begin{equation}
\nu^2 = {9 \over 4} + {3\lambda \over 2} + {\lambda^2 \over 4} + \left( 3 + {5\lambda \over 2} + {\lambda^2 \over 2} \right) \; e^{- \lambda N}.
\end{equation}
and the scalar spectral index is defined through the parameter $\nu$ as in Eq.(\ref{perparam1}). It is realized that the parameter only depends on the second slow-roll parameter $\lambda$ and the number of e-fold $N$, which is required to be around $65$ to solve the problem of hot big bang model. Fig.\ref{NCnslambda} shows the behavior of the scalar spectral index versus $\lambda$ for different choice of the number of e-fold. It is clearly realized that for interested value of the number of e-fold, the estimated value of the model for $n_s$ could not stand  near the observational data. Another point is that, this argue is independent of the form of  $\sqrt{\mathcal{L}_{,X}}$.

%%%%%%%%%%%%%%%%%%%%%%%%%%%%%%%%%%%
%%%%%%%%%%%%%%%%%%%%%%%%%%%%%%%%%%%

%%%%%%%%%%%%%%%%%%%%%%%%%%%%%%%%%%%%%%%%
%%%%%%%%%%%%%%%%%%%%%%%%%%%%%%%%%%%%%
%%%%%%%%%%%%%%%%%%%%%%%%%%%%%%%%%%%%%%%%%%
%%%%%%%%%%%%%%%%%%%%%%%%%%%%%%%%%%%%%%%%
%%%%%%%%%%%%%%%%%%%%%%%%%%%%%%%%%%%%%
%%%%%%%%%%%%%%%%%%%%%%%%%%%%%%%%%%%%%%%%%%%%%
\subsection{Tachyon scalar field model}
\begin{figure}[ht]
    \centering
    \includegraphics[width=6cm]{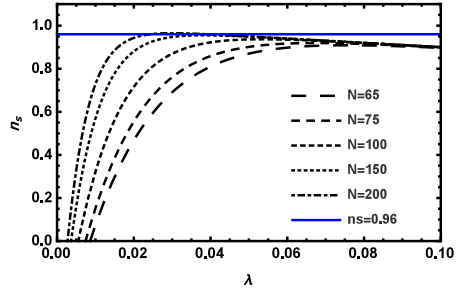}
    \caption{The scalar spectral index versus $\lambda$ for different choice of the number of e-fold has been plotted in the tachyon model.}
    \label{tacns}
\end{figure}

For the next type of k-essence model, here we are going to study the tachyon scalar field model. For the tachyon model, the sound speed is obtained in terms of the slow-roll parameters $\epsilon_1$
\begin{equation}\label{tacsound}
  c_s = \sqrt{1 - {2 \over 3} \; \epsilon_1} \; .
\end{equation}
Then, one could obtain the slow-roll parameter $s$ and its time derivative as
\begin{eqnarray}\label{stachyon01}
    s & = & {-\lambda \over 3} \; {e^{-\lambda N} \over 1 - {2 \over 3}e^{-\lambda N}}, \nonumber \\
    {\dot{s} \over H} & = & \lambda s = {-\lambda^2 \over 3} \; {e^{-\lambda N} \over 1 - {2 \over 3}e^{-\lambda N}}.
\end{eqnarray}
Substituting these quantities in Eq.(\ref{perparam1}), the parameter $\nu$ is found out as
\begin{equation}\label{nutachyon01}
    \nu^2 = {9 \over 4} + {3\lambda \over 2} + {\lambda^2 \over 4} + \left( 3 + {5\lambda \over 2} + {\lambda^2 \over 2} \right) \; e^{-\lambda N}
    + {\lambda (3+2\lambda) \over 3} \; {e^{-\lambda N} \over 1 - {2 \over 3}e^{-\lambda N}}.
\end{equation}
As it is seen, the scalar spectral index for tachyon model only depends on $\lambda$ and number of e-fold $N$. Same as the previous case, the scalar spectral index is extracted in a general form, namely specific function of $\beta$ is not required. Behavior of the scalar spectral index versus second slow-roll parameter $\lambda$ is illustrated in Fig.\ref{tacns} for different choice of number of e-fold. For proper values of number of e-fold, i.e. $55-65$, the scalar spectral index is not compatible with observational data. Then, we are going to leave this case as well. \\
%%%%%%%%%%%%%%%%%%%%%%%%%%%%%%%%%%%
%%%%%%%%%%%%%%%%%%%%%%%%%%%%%%%%%%%

%%%%%%%%%%%%%%%%%%%%%%%%%%%%%%%%%%%%%%%%
%%%%%%%%%%%%%%%%%%%%%%%%%%%%%%%%%%%%%
%%%%%%%%%%%%%%%%%%%%%%%%%%%%%%%%%%%%%%%%%%
%%%%%%%%%%%%%%%%%%%%%%%%%%%%%%%%%%%%%%%%
%%%%%%%%%%%%%%%%%%%%%%%%%%%%%%%%%%%%%
%%%%%%%%%%%%%%%%%%%%%%%%%%%%%%%%%%%%%%%%%%%%%
\subsection{DBI model}
From the dynamical equation of DBI model, it could be concluded that the sound speed for DBI model could be stated as $c_s = -\dot\phi / 2H$, and by using Eq.(\ref{beta}), it could be acquired as
\begin{equation}\label{dbisound}
c_s = \mathcal{L}^{-1}_{,X}.
\end{equation}
which is in agreement with Eq.\eqref{gammaLx} \and \eqref{dbics}. Clearly, the form of the sound speed and its related slow-roll parameters, $s$ and $\dot{s}/H$, depend on the form of the Lagrangian, and the function which is obtained for $\beta$. Therefore, we are going to complete the work, for three choices of $\mathcal{L}_{,X}$ that were made in Sec.\ref{seciii:constantroll}. \\
Before going to more detail about the sound speed and its related slow-roll parameter, there are more result one could get in general. Based on the form of the sound speed it is expected to see the other model constants (rather than $\lambda$ and $N$) in the scalar spectral index, and tensor-to-scala ratio. Therefore, we are going to determined these parameters based on other data that we have about the amplitude of scalar perturbation and the energy scale of inflation [planck]. Based on the amplitude of the scalar perturbation presented in Eq.(\ref{perparam1}), the constant $W_0$ could be found as
\begin{equation}\label{W0}
  W_0 = \varepsilon \left[ 8\pi^2 \; \left( \Gamma(3/2) \over {2^{\nu^\star - {3 \over 2}} \; \Gamma(\nu^\star) } \right)^2 \; e^{-\lambda N} \mathcal{P}^\star_s \; \exp\left( {2 e^{-\lambda N} \over \lambda} \right) \right]^{1 \over 2} \; {1 \over Z(\beta^\star)},
\end{equation}
where $\varepsilon = \pm 1$, and the function $Z(\beta)$ is a typical function chosen as $Z(\beta) \equiv \sqrt{\mathcal{L}_{,X}}$ where the superscript $\star$ indicates that the parameters is evaluated at the horizon crossing time. \\

On the other hand, based on the dynamical equation that were derived in Sec.\ref{kessencemodel}, the potential of the DBI scalar field could be expressed as
\begin{equation}\label{potphi}
  V(\phi) = {3 \over 4} \; W^2(\phi) \; \left[ 1 - {1 \over 3} \; {\left[ \; \beta(\phi) \; Z\big(\beta(\phi)\big) \; \right]^2 \over Z^2\big(\beta(\phi)\big)+1} \right],
\end{equation}
and at the time of horizon crossing, one arrives at
\begin{equation}\label{potstar}
  V^\star = {3 \over 4} \; W_0^2 \exp\left( {-2  \over \lambda} \; e^{-\lambda N} \right) \; \left[ 1 - {1 \over 3} \; {\left( \sqrt{2}\; e^{-\lambda N \over 2} \; Z^\star \right)^2 \over Z^{\star 2}+1} \right],
\end{equation}
where with attention to the choices of $Z(\phi)$ that we made in Sec.\ref{seciii:constantroll}, it is realized that the constant $A_i$ ($i$ is related to the number of typical example) is hidden in $Z^\star$. This constant could be extracted from $Z^\star$ as follow:
\begin{eqnarray}
  Z_1^\star &=& A_1 \tilde{Z}_1^\star, \qquad {\rm where} \qquad \tilde{Z}_1^\star = \sqrt{2^n} \; e^{-n \lambda N \over 2}, \\
  Z_2^\star &=& A_2 \tilde{Z}_2^\star, \qquad {\rm where} \qquad \tilde{Z}_2^\star = \exp\left( \sqrt{2} \; M_2 \; e^{-\lambda N \over 2} \right), \\
  Z_3^\star &=& A_3 \tilde{Z}_3^\star, \qquad {\rm where} \qquad \tilde{Z}_3^\star = {1 \over \cosh\left( \sqrt{2} \; M_2 \; e^{-\lambda N \over 2} \right)}.
\end{eqnarray}
Then for the constant $W_0$, there is
\begin{equation}\label{W0star}
  W_0 = {Q^\star \over A_i},
\end{equation}
where
\begin{equation*}
  Q^\star \equiv \varepsilon \left[ 8\pi^2 \; \left( \Gamma(3/2) \over {2^{\nu^\star - {3 \over 2}} \; \Gamma(\nu^\star) } \right)^2 \; e^{-\lambda N} \mathcal{P}^\star_s \; \exp\left( {2 e^{-\lambda N} \over \lambda} \right) \; \right]^{1 \over 2} \; {1 \over \tilde{Z}_i^\star},
\end{equation*}
and by applying the constant $W_0$ from Eq.(\ref{W0star}) into Eq.(\ref{potstar}), one could determined the constant $A_0$ as
\begin{equation}\label{Ai}
  A_i^2 = {1 \over 2 \xi^\star \tilde{Z}_i^{\star 2} } \left[ -\left( (\beta^{\star 2} - 3)\;\tilde{Z}_i^{\star 2} - \xi^\star \right)
  \pm \sqrt{\left( (\beta^{\star 2} - 3)\;\tilde{Z}_i^{\star 2} - \xi^\star \right)^2 - 4\xi^\star \tilde{Z}_i^{\star 2} }   \right],
\end{equation}
in which the defined quantities are given as
\begin{equation*}
  \xi^\star = {4 V^\star \over Q^{\star 2} \; \exp\left( {-2 \over \lambda} \; e^{-\lambda N}  \right)}.
\end{equation*}
Based on the latest observational data, the amplitude of scalar perturbation is about $\ln\left( 10^{10} \mathcal{P}_s \right) = 2.94$, and for the energy scale of the inflation there is $V^\star < 1.7 \times 10^{-2}$ (note that we put $M_p=1$). Then, the constant $W_0$ and $A_i$ were obtained in terms of the other two constant of the model. In the subsequent lines, we are going to consider the model for specific choice of $\sqrt{\mathcal{L}_{,X}}$, and by using the observational data for the scalar spectral index, and the tensor-to-scalar ratio, determine to remaining constant of the model.

%(******************************************************************)
%(******************************************************************)
%(******************************************************************)
\begin{enumerate}
  \item {\bf Power-Law case:}\\
  For this case, the sound speed of the DBI scalar field is given by
\begin{equation}\label{dbicsa}
    c_s^\star = {1 \over 2^n A_1^2 \; e^{-n\lambda N}}.
\end{equation}
Then, the related slow-roll parameters $s$ and $\dot{s}/H$ are acquired as
\begin{equation}
    s = -n \lambda, \qquad \dot{s} = 0.
\end{equation}
Using above slow-roll parameters in Eq.(\ref{perparam1}), the parameter $\nu$ is obtained as
\begin{equation}
    \nu^2 = {9 \over 4} + {3\lambda \over 2} + {\lambda^2 \over 4} + \left( 3 + {5\lambda \over 2} + {\lambda^2 \over 2} \right) \; e^{-\lambda N}
    + n \lambda (3+\lambda).
\end{equation}
It contribute in scalar spectral index through the relation $n_s=4-2\nu$, which is clearly realized that the scalar spectral index depends of the constant $\lambda$ and $n$, where the number of e-fold is chosen as $N=65$. On the other side, using the sound speed (\ref{dbicsa}), (\ref{Ai}) in (\ref{perparam1}), the tensor-t-scalar ratio is extracted in terms of $\lambda$ and $n$ as well. Then, using the $r-n_s$ diagram of Planck could lead one to obtained the suitable values of the model that put the model prediction
about $n_s$ and $r$ in observational range. Fig.\ref{dbifirstnew} shows this range for the constant.
%%%%%%%%%%%%%%%%%%%%%%%%%%%%%%%%%%%
\begin{figure}[ht]
    \centering
    \includegraphics[width=6cm]{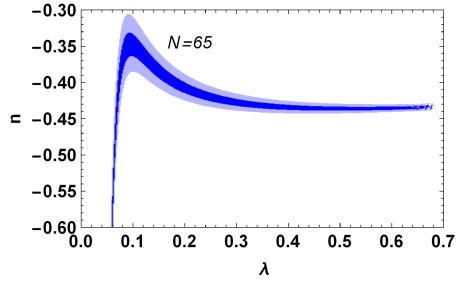}
    \caption{The figures shows the proper values of the model constants $\lambda$ and $n$ where the number of e-fold is $N=65$. }
    \label{dbifirstnew}
\end{figure}
%%%%%%%%%%%%%%%%%%%%%%%%%%%%%%%%%%%
Fig.\ref{dfiepsilon01} shows the behavior of the time derivative of the scalar field and first slow-roll parameter in order to consider the consistency of these two parameter. It is illustrated in Fig.\ref{dfiepsilon01}a that the time derivative of the scalar field is positive during the inflation. Then, passing time and approaching to the end of inflation, the scalar field is increased. On the other hand, $\epsilon=1$ has been assumed as the end of inflation, then it is expected that by passing time and increasing the scalar field, the parameter $\epsilon$ reaches one and indicates the end of inflation. This behaviors is clear from the Fig.\ref{dfiepsilon01} which shows the consistency between the behavior of these two parameters, i.e. $\dot{\phi}$ and $\epsilon$. \\

%%%%%%%%%%%%%%%%%%%%%%%%%%%%%%%%%%%
\begin{figure}[h]
    \centering
    \subfigure[]{\includegraphics[width=6cm]{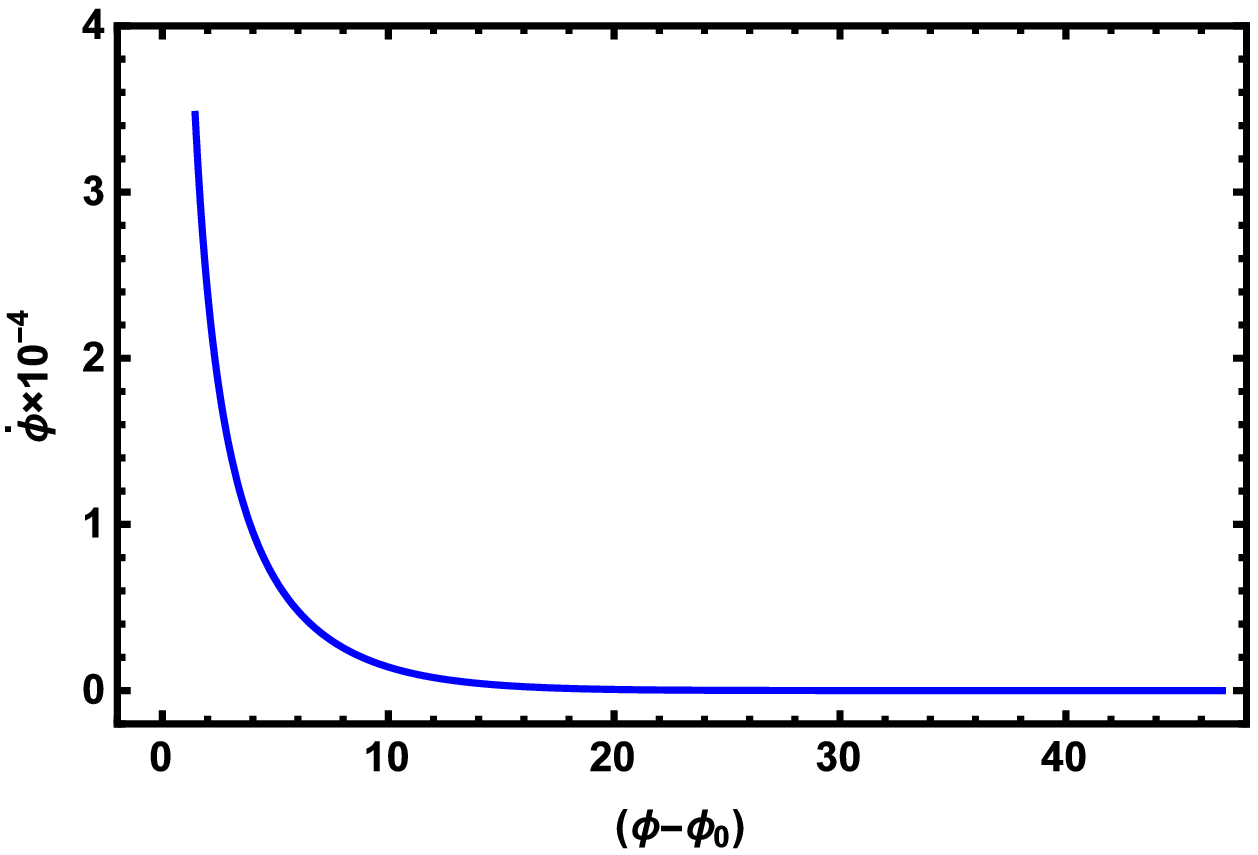}}
    \subfigure[]{\includegraphics[width=6cm]{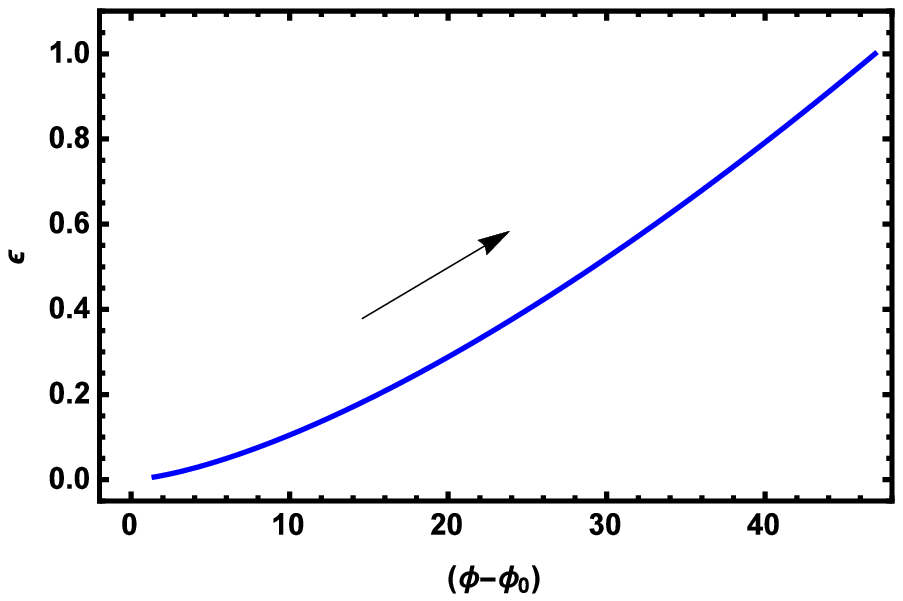}}
    \caption{Behavior of a) $\dot\phi$, and b) $\epsilon_1$ during the inflation versus the scalar field. }
    \label{dfiepsilon01}
\end{figure}
%%%%%%%%%%%%%%%%%%%%%%%%%%%%%%%%%%%
As the final step, the behavior of the potential of the model for this case is presented in Fig.\ref{potDBI01}. Since the time derivative of the scalar field is positive, the scalar field has smaller values at the beginning of inflation. Therefore, the scalar field at the beginning of inflation is at the top of its potential, and by passing time it slowly rolls down to the bottom of its potential, which is clear from the Fig.\ref{potDBI01}. The energy scale of inflation could be also read from the plot, which declares that the energy scale of inflation is about $(V^\star)^{1/4} \propto 10^{-3}(M_p)$ which is smaller than the Planck energy scale.  \\
%%%%%%%%%%%%%%%%%%%%%%%%%%%%%%%%%%%
\begin{figure}
\includegraphics[width=7cm]{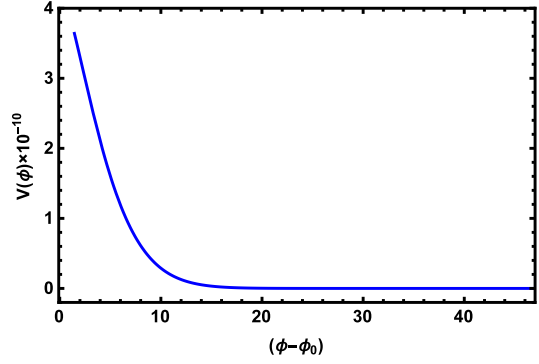}
\caption{Behavior of the potential of the scalar field for DBI model with the first typical choice of the Lagrangian.}\label{potDBI01}
\end{figure}
%%%%%%%%%%%%%%%%%%%%%%%%%%%%%%%%%%%

%%%%%%%%%%%%%%%%%%%%%%%%%%%%%%%%%%%
%%%%%%%%%%%%%%%%%%%%%%%%%%%%%%%%%%%
%%%%%%%%%%%%%%%%%%%%%%%%%%%%%%%%%%%
  \item {\bf Exponential case:}\\
  Using the second introduced ansatz, and by applying results of Sec...., the sound speed and the related slow-roll parameters at the time of horizon crossing are obtained as
\begin{eqnarray}\label{dbicsb}
  c_s &=& {1 \over A_2^2} \; \exp\left( 2\sqrt{2}\; M_2 e^{-\lambda N \over 2} \right), \nonumber \\
  s &=& \sqrt{2} \; \lambda M_2 \; e^{-\lambda N \over 2} , \nonumber \\
  {\dot{s} \over H} &=& {\lambda \over 2} \; s = {\lambda^2 \over \sqrt{2}} \; M_2 \; e^{-\lambda N \over 2}.
\end{eqnarray}
With above relation, the corresponding $\nu$ is given by
\begin{equation}\label{nudbisecond}
  \nu^2 = {9 \over 4} + {3\lambda \over 2} + {\lambda^2 \over 4} + \left( 3 + {5\lambda \over 2} + {\lambda^2 \over 2} \right) \; e^{-\lambda N}
    - \sqrt{2} \; \lambda M_2  (3+\lambda) \; e^{-\lambda N \over 2}
    - {\lambda^2 \over \sqrt{2}} \; M_2 \; e^{-\lambda N \over 2},
\end{equation}
which vividly clarifies that the scalar spectral index depends on the second slow-roll parameter $\lambda$ and $M_2$. To find out the dependence of the tensor-to-scalar ratio on the model constant, one could use the sound speed (\ref{dbicsb}) and (\ref{Ai}) and insert them into Eq.(\ref{perparam1}). It is shown that same as $n_s$, the tensor-to-scalar ratio also depends on $\lambda$ and $M_2$ (where the number of e-fold is picked out as $N=65$). To determine these constant, one can use the Planck $r-n_s$ diagram, and restrict the model constant as has been performed in Fig.\ref{dbisecond}, where the dark and light blue area respectively show the range of the constants that for every point in this range the model result stands in $68\%$ and $95\%$ CL area. \\

%%%%%%%%%%%%%%%%%%%%%%%%%%%%%%%%%%%
\begin{figure}[h]
    \centering
    \includegraphics[width=6cm]{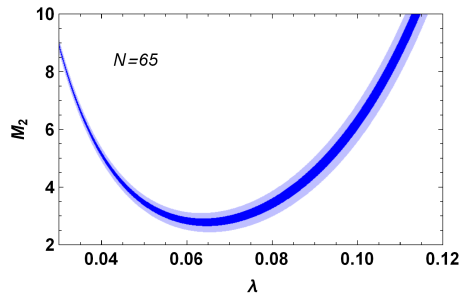}
    \caption{The figures shows the proper values of the model constants $\lambda$ and $M_2$ where the number of e-fold is $N=65$. }
    \label{dbisecond}
\end{figure}
%%%%%%%%%%%%%%%%%%%%%%%%%%%%%%%%%%%
So far we could determine the free parameters of the model using the observational data for the perturbation parameters and energy scale of the inflation based on Planck data. In this step, we are about to use these free parameters and consider the behavior of the time derivative of the scalar field and the slow-roll parameter $\epsilon_1$ during inflation, and investigate their compatibility. Fig.\ref{dfiepsilon02} describe behavior of these two parameters during inflation. It is clearly seen in Fig.\ref{dfiepsilon02}a that $\dot\phi$ is always negative which means that the scalar field decreases during inflation. Then, the scalar field value at the time of horizon crossing should be bigger than the scalar field value at the end of inflation. Consequently, it is expected that the slow-roll parameter $\epsilon_1$ be smaller than unity for bigger values of the scalar field and increases by passing time and reducing of the scalar field. It is confirmed in Fig.\ref{dfiepsilon02}b, where one could find the behavior of $\epsilon_1$ versus scalar field during inflation. \\

%%%%%%%%%%%%%%%%%%%%%%%%%%%%%%%%%%%
\begin{figure}[h]
    \centering
    \subfigure[]{\includegraphics[width=6cm]{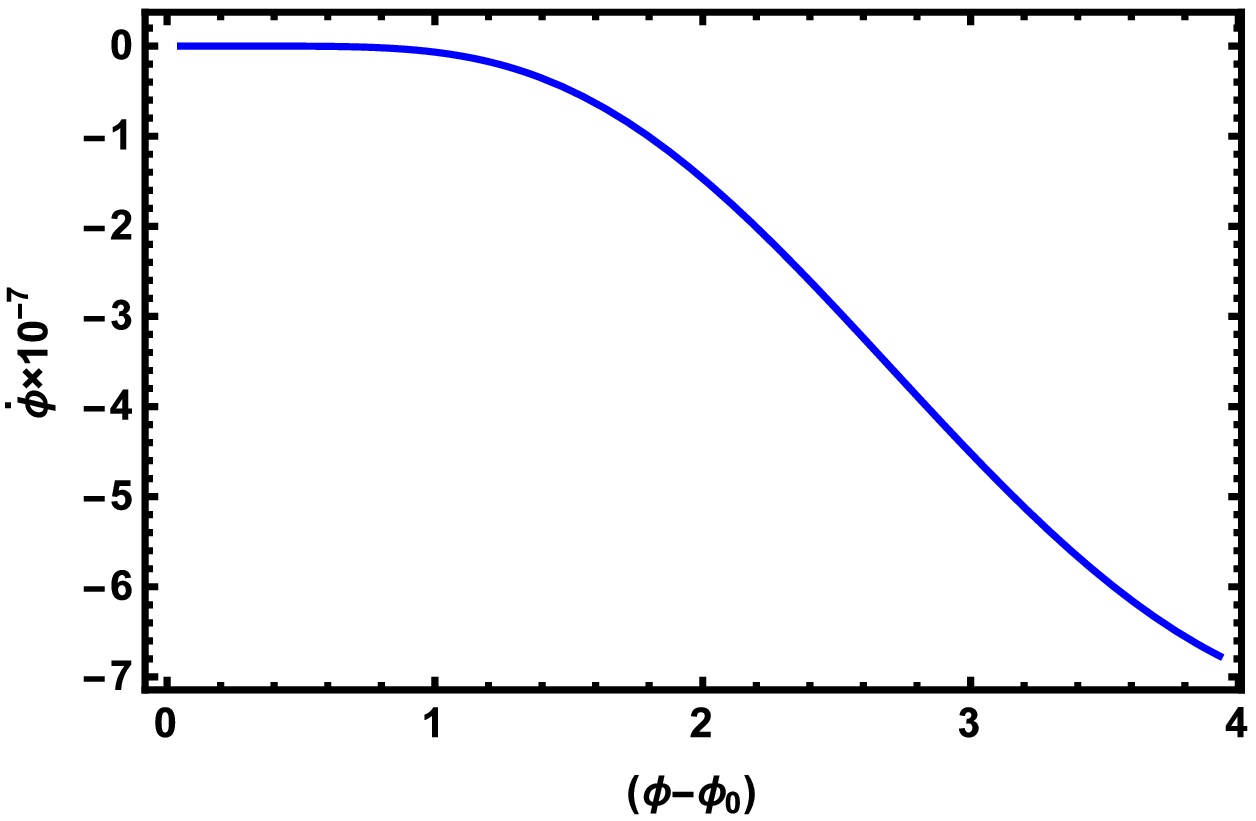}}
    \subfigure[]{\includegraphics[width=6cm]{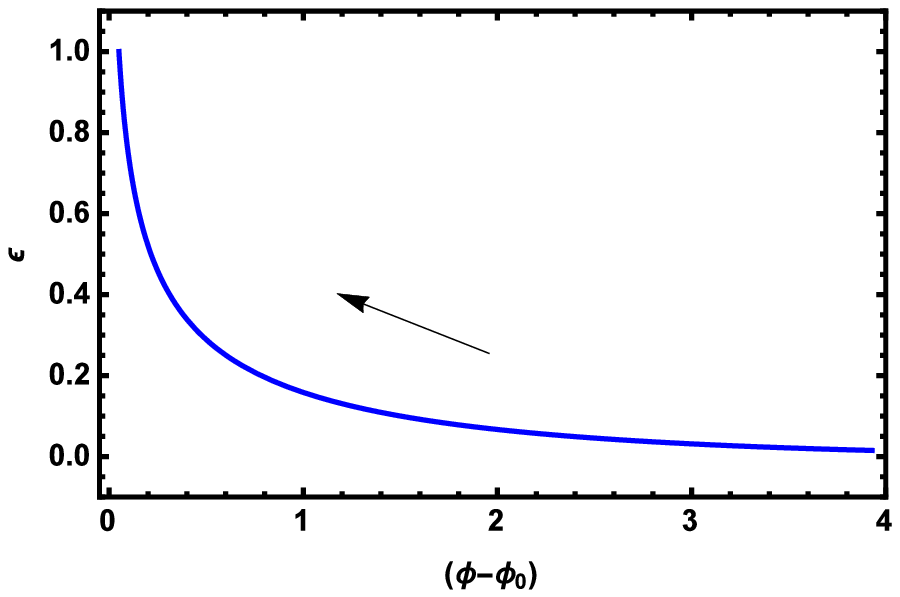}}
    \caption{Behavior of a) $\dot\phi$, and b) $\epsilon_1$ during the inflation versus the scalar field. }
    \label{dfiepsilon02}
\end{figure}
%%%%%%%%%%%%%%%%%%%%%%%%%%%%%%%%%%%

The behavior of the potential during inflation for this case is depicted in Fig.\ref{potDBI02}. As it was discussed the time derivative of the scalar field is negative, then the scalar field at the beginning of inflation has bigger values. Hence, the Fig.\ref{potDBI02} clearly determines that the scalar field stands at the top of its potential at the beginning, then it slowly rolls down to the bottom. The energy scale of inflation for this case is about $(V^\star)^{1/4} \propto 10^{-3}(M_p)$. \\

%%%%%%%%%%%%%%%%%%%%%%%%%%%%%%%%%%%
\begin{figure}[h]
\includegraphics[width=7cm]{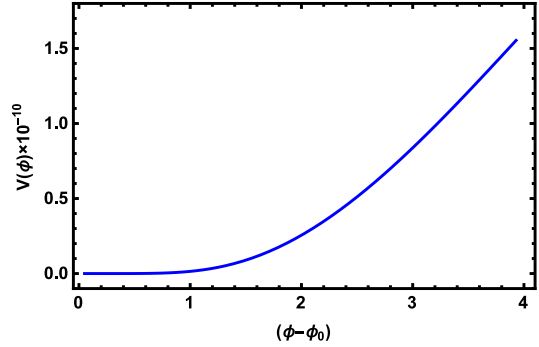}
\caption{Behavior of the potential of the scalar field for DBI model with the second typical choice of the Lagrangian.}\label{potDBI02}
\end{figure}
%%%%%%%%%%%%%%%%%%%%%%%%%%%%%%%%%%%

%%%%%%%%%%%%%%%%%%%%%%%%%%%%%%%%%%%
%%%%%%%%%%%%%%%%%%%%%%%%%%%%%%%%%%%
%%%%%%%%%%%%%%%%%%%%%%%%%%%%%%%%%%%
  \item {\bf Hyperbolic case:}\\
  For the third ansantz, the sound speed of DBI scalar field is read as
\begin{equation}\label{csdbi3}
c_s = {1 \over A_3^2} \; \cosh^2\left( \sqrt{2} \; M_3 e^{-\lambda N} \right).
\end{equation}
and the corresponding slow-roll parameters $s$ and its time derivative come to
\begin{eqnarray}
    s & = & \sqrt{2} \; \lambda \; M_3 e^{-\lambda N \over 2} \tanh\left( \sqrt{2} \; M_3 e^{-\lambda N \over 2} \right), \\
    {\dot{s} \over H} & = & {\lambda s \over 2} \; \left( 1 + {\sqrt{2} \; M_3 e^{-\lambda N \over 2} \over \sinh\left[ \sqrt{2} \; M_3 e^{-\lambda N \over 2} \right] \cosh\left[ \sqrt{2} \; M_3 e^{-\lambda N \over 2} \right]} \right)
\end{eqnarray}
Applying the above conclusion, the parameter $\nu$ is derived as
\begin{eqnarray}
    \nu^2 & = &  {9 \over 4} + {3\lambda \over 2} + {\lambda^2 \over 4}
    + \left( 3 + {5\lambda \over 2} + {\lambda^2 \over 2} \right) \; e^{-\lambda N} \\
     & & \qquad - (3+\lambda) \sqrt{2} \lambda M_3 \; e^{-\lambda N \over 2} \tanh\left[ \sqrt{2} \lambda M_3 \; e^{-\lambda N \over 2} \right] \nonumber \\
    & & \qquad - {\lambda^2 \over 2} \sqrt{2}\;  M_3 \; e^{-\lambda N \over 2} \tanh\left[ \sqrt{2} \lambda M_3 \; e^{-\lambda N \over 2} \right] \; \left( 1 + {\sqrt{2} \; M_3 e^{-\lambda N \over 2} \over \sinh\left[ \sqrt{2} \; M_3 e^{-\lambda N \over 2} \right] \cosh\left[ \sqrt{2} \; M_3 e^{-\lambda N \over 2} \right]} \right). \nonumber
\end{eqnarray}
Through the relation $n_s=4-2\nu$, it is found out that the scalar spectral index is only depends on the second slow-roll parameter $\lambda$ and $M_3$. In addition, by imposing Eqs.(\ref{dbicsc}) and (\ref{Ai}) on Eq.(\ref{perparam1}), the tensor-to-scalar ratio is extracted also in terms of $\lambda$ and $M_3$. To constraint the constants, the Planck $r-n_s$ diagram is used and one could find a range for every constant which put the model in agreement with observational data. Fig.\ref{dbithird} portrays these ranges for the constants $\lambda$ and $M_3$ in which the light and dark blue range indicate values of the constants that the result stands respectively in $95\%$ and $68\%$ CL. \\
%%%%%%%%%%%%%%%%%%%%%%%%%%%%%%%%%%%
\begin{figure}[ht]
    \centering
    \includegraphics[width=6cm]{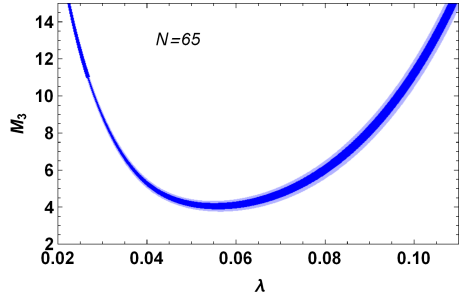}
    \caption{The figures shows the proper values of the model constants $\lambda$, $M_3$, when one picks out the third ansantz for the term $\mathcal{L}_{,X}$, and DBI Lagrangian. The result is plotted for number of e-fold $N=65$.  }
    \label{dbithird}
\end{figure}
%%%%%%%%%%%%%%%%%%%%%%%%%%%%%%%%%%%
Following above approach, we could specify the free parameters of the model by applying the latest observational data of Planck. By applying these results, one could consider the behavior of the time derivative of the scalar field and first slow-roll $\epsilon_1$ during inflation and check the consistency of these two parameter. In this regards, Fig.\ref{dfiepsilon03}a depicts $\dot\phi$ and $\epsilon_1$ in terms of the scalar field. In contrast to the previous case, here $\dot\phi$ is positive during the whole inflationary times which means the scalar field increase by passing time, and at the end of inflation we have a bigger scalar field value than the initial scalar field value. Therefore, the slow-roll $\epsilon_1$ is expected to be smaller than unity at smaller values of the scalar field and by passing time and growing scalar field it should increase and reaches one. This expectation is confirmed in Fig.\ref{dfiepsilon03}b. \\
%%%%%%%%%%%%%%%%%%%%%%%%%%%%%%%%%%%
\begin{figure}[h]
    \centering
    \subfigure[]{\includegraphics[width=6cm]{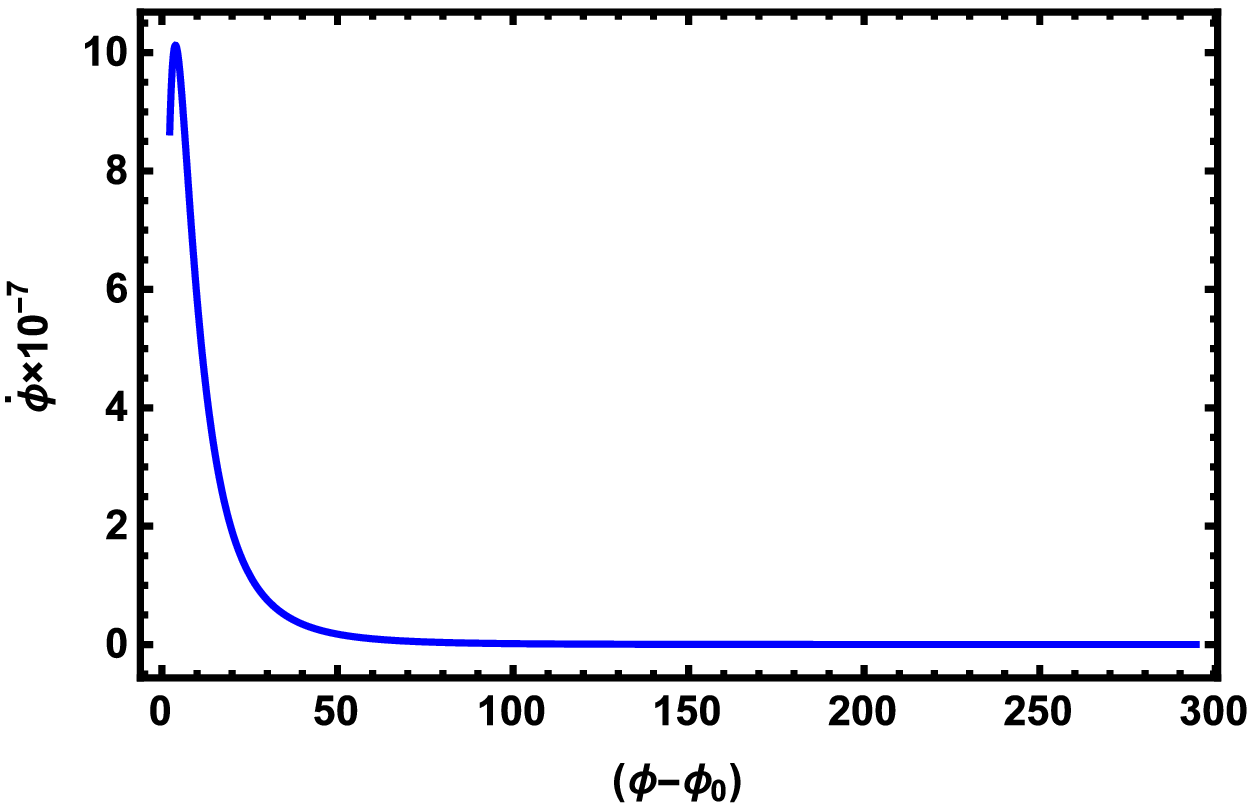}}
    \subfigure[]{\includegraphics[width=6cm]{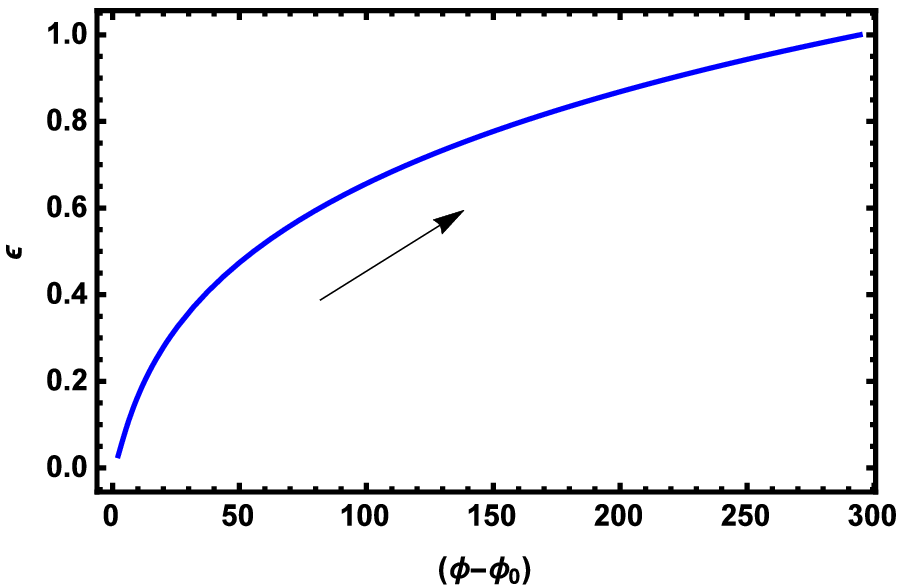}}
    \caption{Behavior of a) $\dot\phi$, and b) $\epsilon_1$ during the inflation versus the scalar field. }
    \label{dfiepsilon03}
\end{figure}
%%%%%%%%%%%%%%%%%%%%%%%%%%%%%%%%%%%

The potential of the scalar field for this case is illustrated in Fig.\ref{potDBI03}, where the inflation begins for the smaller values of the scalar field (due to the positiveness of the $\dot{\phi}$). At the beginning, the scalar field stands on the top of its potential, then it rolls down to the bottom. The energy scale of inflation is as the same order as the previous cases as $(V^\star)^{1/4} \propto 10^{-3}(M_p)$. \\

%%%%%%%%%%%%%%%%%%%%%%%%%%%%%%%%%%%
\begin{figure}[h]
\includegraphics[width=7cm]{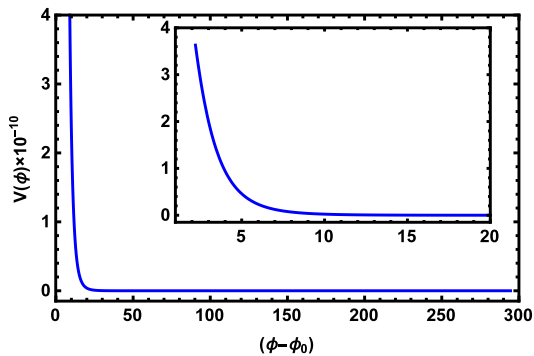}
\caption{Behavior of the potential of the scalar field for DBI model with the third typical choice of the Lagrangian.}\label{potDBI03}
\end{figure}
%%%%%%%%%%%%%%%%%%%%%%%%%%%%%%%%%%%

\end{enumerate}

%(******************************************************************)
%(******************************************************************)
%(******************************************************************)

%%%%%%%%%%%%%%%%%%%%%%%%%%%%%%%%%%%%%%%%
%%%%%%%%%%%%%%%%%%%%%%%%%%%%%%%%%%%%%
%%%%%%%%%%%%%%%%%%%%%%%%%%%%%%%%%%%%%%%%%%%%%
%%%%%%%%%%%%%%%%%%%%%%%%%%%%%%%%%%%%%%%%%%%%%
%%%%%%%%%%%%%%%%%%%%%%%%%%%%%%%%%%%%%%%%
%%%%%%%%%%%%%%%%%%%%%%%%%%%%%%%%%%%%%
%%%%%%%%%%%%%%%%%%%%%%%%%%%%%%%%%%%%%%%%%%%%%
\section{Conclusion}\label{conclusion}
Utilizing $\beta$-function formalism, the constant-roll inflationary scenario for k-essence model of cosmology was considered. The Lagrangian of the k-essence model is assumed to be a function of the scalar field $\phi$, and its time derivative $X$. The k-essence model is known as a wide model in which many known cosmological models such as non-canonical, tachyon and DBI are recognized as a subclass of it. Inspiring from renormalization group equation, the recently proposed approach named as $\beta$-function formalism seems to be eligible method for considering constant-roll inflationary scenario, especially when the constancy of the second slow-roll parameter results in first order differential equation; in contrast to the usual approach that gives a second order differential equation which could be also non-linear. However, for the case of k-essence model, the term $\mathcal{L}_{,X}$ appears in the differential equation of $\beta$, which makes it difficult to find a solution in general. Regardless of solving the equation and finding the $\beta$ in terms of the scalar field, $\beta$-function, the slow-roll parameter $\epsilon$ and the superpotential are determined at the time of horizon crossing and at the end of inflation. The first slow-roll parameter was acquired as $\epsilon=e^{-\lambda N}$ which states that the second slow-roll parameter should be positive to arrives at a small value or $\epsilon$ at the time of horizon crossing which is our fundamental assumption. However in order to go through more detail, it was assumed that the term $\mathcal{L}_{,X}$ could be stated in terms of $\beta$ and a general solution was found for three different choice of $\mathcal{L}_{,X}$.\\
Cosmological perturbations of the model was investigated, and the important parameter $z$ was expressed in terms of the $\beta$-function, scale factor and the sound speed which depends on the type of the chosen Lagrangian. The parameter $z$ and its derivative have fundamental role in the dynamical equation of the scalar perturbations, and appears in the main perturbation parameters such as curvature perturbations, scalar spectral index, and tensor-to-scalar ratio.    \\
Considering the compatibility of the model with observational data was followed in next step, and constraining the free parameter by using the data was sought. In this regards, three well-known Lagrangian were picked out as non-canonical, tachyon and DBI models, and for each case the three cases of the mentioned ansatz were investigated. For non-canonical model it is found out that the scalar spectral index could not reach the observational range for the interested values of the number of e-fold. As a matter of fact, for any model with constant sound speed, the scalar spectral index could not reach the observational range. For the second model, the tachyon scalar field model was considered, and despite the fact that for tachyon model the sound speed in varying the scalar spectral index meet the observational data for the scalar spectral index only for high values of the number of e-fold which is not acceptable since we are looking for the number of e-fold about $55-65$. Last Lagrangian belongs to the DBI scalar field model. Using the observational data for energy scale of the universe and the curvature perturbation, the two free parameters of the model was determined. Then, by using the Planck $r-n_s$ diagram it was aimed to plot the acceptable range for other free parameters in which to get an agreement with data. In this regards, the situation was considered for each ansatz. The results indicates that if the term $\mathcal{L}_{,X}$ is taken as a power-law function of the $\beta$, then the model could not achieved the goal, however for other two cases, it is possible to constrain the parameter so that to have our interest values for the perturbation parameters and reach to an agreement between the model prediction and observational data. \\
The time derivatives of the scalar field reveals whether the scalar field increases or decreases during the time. On the other hand, since the end of inflation is taken as $\epsilon=1$, the consistency of these two parameter should be noted in which if $\dot\phi<0 (>0)$ then $\epsilon$ should reached unity for smaller (larger) values of the scalar field. To check this consistency, the behavior of the time derivatives of the scalar field and the first slow-roll parameter was plotted separately for the last two cases of DBI model by using the free parameters that have just determined. The Figs.\ref{dfiepsilon02} and \ref{dfiepsilon03} clearly portray that by having a negative $\dot\phi$ the slow-roll parameter $\epsilon$ reaches one for the smaller values of the scalar field and for positive $\dot\phi$ the slow-roll parameter $\epsilon$ arrives at one for bigger values of the scalar field. \\
The potential of the scalar field for all cases is studied as well, where it is shown that, as it was expected, at the beginning of inflation the scalar field stands on the top of its potential, then it rolls down to the bottom. The energy scale of the inflation was about $(V^\star)^{1/4} \propto 10^{-3}(M_p)$ which is smaller than the Planck energy scale.

\nocite{*}

\bibliography{Refbib}% Produces the bibliography via BibTeX.

\end{document}